\documentclass[final,5p,times,twocolumn]{elsarticle}
\usepackage{float,subcaption,lineno,hyperref,sidecap,paralist,dblfloatfix,booktabs,multicol}
\usepackage{makecell}  
\usepackage{siunitx}

\sidecaptionvpos{figure}{t}
\modulolinenumbers[5]
\begin{document}
\begin{frontmatter}

\title{Microstructure Effects on Performance and Deactivation of Hierarchically Structured Porous Catalyst: a Pore Network Model}

\author[AUT]{Masood Moghaddam}
\ead{m.moghaddam@aut.ac.ir}

\author[AUT]{Abbas Abbassi\corref{cor1}}
\cortext[cor1]{Corresponding author}
\ead{abbassi@aut.ac.ir}
\author[ZNU]{Jafar Ghazanfarian}
\ead{j.ghazanfarian@znu.ac.ir}

\address[AUT]{Department of Mechanical Engineering, Amirkabir University of Technology (Tehran Polytechnic), 424 Hafez Ave., P.O. Box 1591634311, Tehran, Iran}
\address[ZNU]{Faculty of Engineering, Department of Mechanical Engineering, University of Zanjan, University Boulevard 45371-38791 Zanjan, Iran}

\journal{Chemical Engineering Journal}

\begin{abstract}
In this paper, the pore network model to investigate the reaction-diffusion process in the hierarchically structured porous catalyst particle is extended to consider the phenomenon of deactivation by coking. In this framework, the interaction of internal particle pore structure and mass transfer under the condition of coke deposition are examined. A primitive experimental investigation has been performed as an introduction to the development of the model. Then, the effect of structural features namely macroporosity and pore size ratio, the deactivation properties, the maximum loading of coke as well as the transport properties, the pore Damk\"ohler number on the net reaction rate and deactivation of the particle have been investigated. Three deactivation mechanisms are accounted for, namely, the site coverage, the pore narrowing, and the pore blockage. It is found that the deactivation of the catalyst particle can be divided into two conditions: the kineticsal deactivation and the structural deactivation. It is shown that depending on the Damk\"ohler number, increasing the macroporosity does not necessarily improve the reactivity and deactivation resistance of the catalyst. The key finding of this work is to demonstrate and quantify how changing the typical fresh catalyst microstructure observed in the experimental characterization into a hierarchical one influences the reactivity and deactivation.
\end{abstract}


\end{frontmatter}

\section{Introduction}
\label{Introduction}
A unimodal pore size distribution ranging in the nanometer scale provides high surface area for the porous materials, but their transport properties are expected to be weak due to hindered diffusivity in nanopores. To improve mass transport characteristics in nanoporous materials, the utilization of hierarchical structures is an area of growing interest. Hierarchically structured porous materials are nature inspired structures in which the pore structure is engineered such that the resulting material has desired structure-function properties. For example, in the context of fuel cells, outstanding mass transport properties, high active reaction sites and mechanical strength are three features of an ideal electrode that can be provided by three-dimensional hierarchical porous structures.

In the context of catalysis, high surface area and outstanding transport properties simultaneously are two parameters of an ideal catalyst support; and hierarchical porous catalysts have been proved to be a promising solution to provide both features. By development of new controllable technologies for catalyst synthesis, understanding the effect of pore structure on performance has become more important than ever. Pore-scale models that consider the real geometrical heterogeneity for obtaining detailed pore-scale information of mass transport and reaction have attracted significant attention in recent years \cite{sadeghi2017pore,chen2018pore,moghaddam2020investigation}.

Pore network modeling is one of the best pore-scale approaches for the investigation of transport in porous media, specially in materials with hierarchical structures \cite{dashtian2019convection,vovcka2000pore}. The study of Petropoulos et al. \cite{petropoulos1991network} is a groundwork in the field of transport in hierarchical materials which is implemented to determine the relationship between the gas transport properties and the pore structure. Their framework was extended by Meyers and Liapis \cite{meyers1998network,meyers1999network} to take into account the effects of pore connectivity. Such studies which use the pore network modeling to investigate the transport in hierarchical materials date back almost two decades and consequently an extensive study on the interplay between structural features and transport properties in the context of hierarchical materials is missing in the literature. As an attempt to fill this gap, Sadeghi et al. \cite{sadeghi2017pore} presented a general framework based on the pore network modeling for the simulation of reactive transport in a porous catalyst with a hierarchy of porosity. They also performed a parametric study to investigate the effect of porous structures on particle performance. They showed that particles with lower macropore to nanopore diameter ratio are always more kinetically active, but particles with higher macroporosity are not.

More recently, Moghaddam et al. \cite{moghaddam2020investigation} developed a powerful framework based on the pore network model to investigate the reactive transport phenomena in hierarchically structured porous catalysts. Results showed the benefits of the hierarchical structures created from the existence of macropores inside the nanoporous network as well as their interconnections. Depending on the different reaction conditions (i.e. the average pore Damk\"ohler numbers), three distinct trends were observed for the particle performance by macroporosity increment, including the monotonic decrease, the mixed increase-decrease, and the monotonic increase at low, moderate, and high average pore Damk\"ohler numbers, respectively. So, the key funding of this study was that the benefit of using hierarchical structure is restricted to diffusion limited processes where the diffusivity of species becomes a limiting factor for the reactive transport. In such a process, the reactant is totally consumed near the particle boundary, and most of the particles are not accessible to the reactant. In this case, adding macropores can significantly improve the diffusivity, and the total reaction rate monotonously increases by increasing the macroporosity. This trend also has been confirmed by the results obtained by the lattice Boltzmann method \cite{chen2018pore}.

Higher resistance to deactivation is another important feature expected from an efficient catalyst. Deactivation of catalysts is a ubiquitous problem that causes loss of catalytic rate with time. Coke deposition, which can lead to coverage of active sites and blockage of pores has been proved to be one of the most important reasons for catalyst deactivation \cite{wang2021pore}. The design of active sites and optimization of the pore structure of the catalyst are two methods of resolving deactivation problems. Numerous studies have been conducted on catalyst active site design (leading to significant  progress to improve the catalyst coke resistance), but few studies have focused on the catalyst pore structure so far \cite{lin2021lattice}. This point receives less attention in comparison to designing the catalytic active sites \cite{ye2019optimizing}. Rao and Coppens \cite{rao2012increasing} showed that hydrodemetalation catalyst with optimized pore structure can achieve nearly double useful lifetime as well as higher robustness against the catalyst deactivation. Schmidt et al. \cite{schmidt2013coke} investigated ZSM-5 with a hierarchical pore structure and found that it has a higher catalyst lifetime and methanol conversion capacity in methanol to hydrocarbon process in comparison to conventional ZSM-5. Lin et al. \cite{lin2021lattice} investigated the hierarchical structure-performance relationship in dry reforming of methane in order to further enhance the catalytic activity and inhibit carbon formation. For this purpose, they studied the effects of three hierarchical pore geometrical parameters, namely the catalyst porosity, the ratio of mesopore volume to macropore volume, and the ratio of average macropore diameter to average mesopore diameter on coke formation and catalytic performance to elucidate the deactivation and reaction-diffusion mechanism of the catalyst.

So it can be concluded that the problem of deactivation by coking is highly affected by the internal structure of the catalyst, and using hierarchical structures can improve the resistance to deactivation. But, although the pore network modeling is a powerful tool for studying the transport phenomena in hierarchical porous catalysts, little attention has been paid to using the pore network modeling for this purpose. To the best of our knowledge, the only use of pore network model for studying deactivation problem of hierarchical catalyst is the work of Ye et al. \cite{ye2019optimizing} in which a pore network model has been developed to study the coke formation for propane dehydrogenation. It is found that the optimized hierarchical pore structure resulted in $14$ times higher activity. Although interesting, the size of the network in this study is inclusive of only $ 2139 $ pores which is too small in comparison to $ 33 $ million pores of the hierarchical network modeled with the framework presented in our recent paper \cite{moghaddam2020investigation}. In addition, this study lacks generality and is limited to the investigation of the effect of porosity and pore connectivity on the particle performance. Hence, this study does not provide a clear outcome of the interplay between structural features of the hierarchical particles and the transport properties. So, a comprehensive study on the interplay between the structural parameters and the transport properties in the context of activation and deactivation of hierarchical catalysts is missing in the literature. Thus, it would be very valuable if the pore network based framework presented in our recent paper \cite{moghaddam2020investigation} can be extended to include the deactivation phenomenon. Such a framework can be used for estimating performance in terms of the net reaction rate of an arbitrary hierarchical porous catalyst particle in dynamic conditions that emerged due to deactivation. In this situation, the catalyst pore network structure affects the deactivation by coking, and in turn, the possible coking effects like the pore blocking or narrowing gradually can alert the pore network structure.

The current work has been performed on a hierarchically porous catalyst particle with a bidisperse pore size distribution, such as interlacing macropores in nanoporous particles by the use of porogens or pore-formers. The aim of this work is to facilitate designing hierarchical catalyst supports and for this purpose, the hierarchical particles with certain structural features as tunable parameters are realized in \textit{silico}. Then, by performing a parametric study on the tunable parameters, the performance curves in terms of the net reaction rate within the particle and the dimensionless coke content curves are obtained, which can be considered as guidelines for designers of catalyst supports.

Catalyst deactivation due to coke formation can be divided in to three mechanisms: the site coverage, the pore narrowing, and the pore blockage \cite{froment2001modeling}. In the site coverage, the coke covers the catalytic active sites in which they are no longer accessible for reactants. In the pore narrowing, with formation of coke on the surface of pores, the coke narrows the pores, and consequently increases the diffusion resistance that affects the catalyst activity. Finally, in the pore blockage, the coke plugs pores which means that the pores have no longer any contribution in the reaction and diffusion. When the pore blockage occurs in the outer pores, it may leads to creation of inaccessible clusters of open pores and as a result, dropping the effective diffusivities of the components and the observed reaction rates \cite{arbabi1991computer,ye2019optimizing}. In this study, all these three mechanisms are considered for the investigation of the catalyst deactivation.
Finally, it should be mentioned that a parametric sweep of a variety of particle physical properties is considered for presenting the results of this study in order to understand the effects of different structures on the activity and coke resistance properties of the nanoparticle. The aim of the current study is to determine a recipe that provides the highest activity as well as the coke resistance with the smallest amount of precious catalyst material. Therefore, a first-order reaction kinetics were assumed for this study. However, nothing changes in the case of nonlinear reaction kinetics except that the system of equations should be solved iteratively.


\section{Model development}
Although catalyst deactivation by coking is inevitable, some of its effects may be prevented and delayed. Catalyst microstructure, spatially structured hierarchy, can influence the reactivity and the coke resistance of porous catalyst particles. In order to study this structure-function problem, an appropriate pore scale model should be employed. Inadequacy of continuum models for such pore scale studies are discussed in \cite{sadeghi2017pore} and \cite{moghaddam2020investigation}. The pore network model is one of the best choices for investigating reactive transport in the presence of the coking phenomenon in a porous catalyst with a hierarchy of porosity. This is because, the hierarchical structure of the pore network as well as the pore narrowing and blocking due to coke formation can be considered during the network generation. The pore network model-based framework presented in our recent paper \cite{moghaddam2020investigation} is extended here in order to include the phenomena of deactivation by coking.

\subsection{Experimental characterization}
\label{EXP}
\begin{figure*}[t!]
	\centering
	\includegraphics[width=\linewidth]{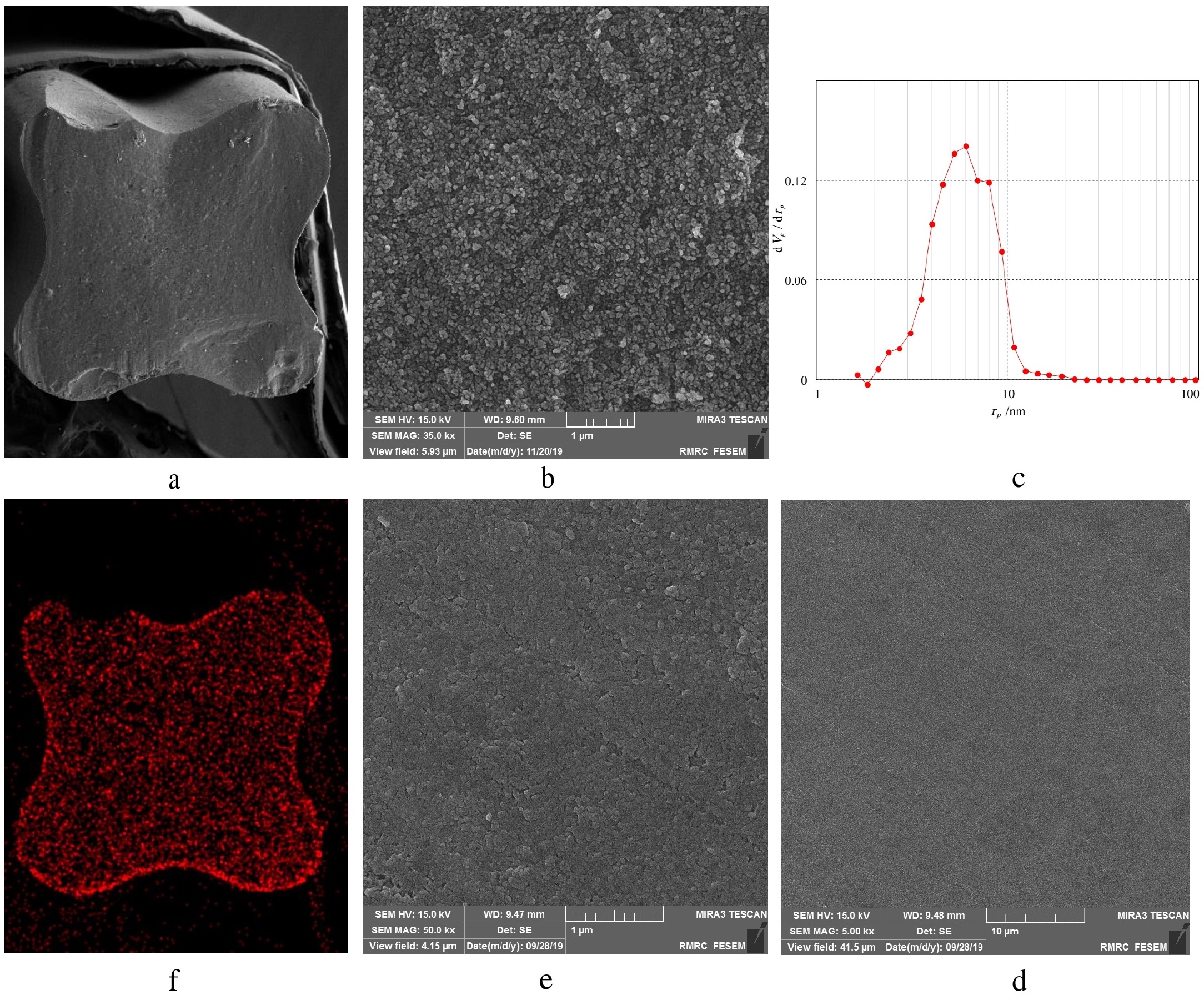}%
	\caption{Results of characterizing fresh HDS catalyst by FESEM (a, b, d and e), element mapping (f) and the BJH analysis (c).}
	\label{Fresh}
\end{figure*}

Due to the  deactivation of the catalyst, its effectiveness changes with time. Coking is one of the deactivation causes, which occurs in a wide range of commercialized processes, and leads to major losses in the industry worldwide every year. For example, the $ Pt-Sn/ZnAl_{2}O_{4} $ catalyst of the Uhde STAR process needs to be regenerated after $ 7 $ hr on stream and the $ CrO_{x}/Al_{2}O_{3} $ catalyst of the Catofin process is deactivated after around $ 12 $ min  of operation \cite{ye2019optimizing}. Fig.~\ref{Fresh}-a and b show the cross section of a fresh commercial hydrodesulfurization (HDS) catalyst which was obtained by the field emission scanning electron microscopy (FESEM) using TESCAN MIRA3 FESEM. In this sample, like most industrial catalysts, the pore size distribution in sub-micron scale is uniform (and unimodal). Fig.~\ref{Fresh}-c illustrates the pore diameter distribution calculated from  the desorption isotherms measured by Belsorp adsorption-desorption (BEL Japan Inc.) using the Barrett-Joyner-Halenda (BJH) method. Uniform pore size distribution shown in this analysis can be better understood by polishing the cross section. For this purpose, the cross section of the catalyst in Fig.~\ref{Fresh}-a is well-polished and then recharacterized by FESEM. Fig.~\ref{Fresh}-d and e demonstrates the cross section of the polished sample which is completely nanoporous and there is no trace of macropores. In addition to uniform pore size distribution, industrial catalysts have uniform distribution of  active sites. The FESEM chemical mapping image of the catalyst cross section is presented in Fig.~\ref{Fresh}-f that definitely shows the homogeneous distribution of Molybdenum (Mo), which is the main active element of the HDS catalysts.

In the catalytic process, the phenomenon of deactivation by coking leads to covering the active sites and blocking the pores. This means that the initial distribution of the pore size and accessibility of the active sites in the fresh catalyst will be changed over time. In order to extend our previews pore network model \cite{moghaddam2020investigation} to investigate the structure-function relation in presence of the coking phenomenon, the pore narrowing (and maybe blocking) and active site covering should be considered in the development of the model. For this purpose, the initial pore network should be updated at each time-step in order to consider the effect of pore narrowing by coke deposition and possible pore blockage. Also, the active site coverage should be considered in the mathematical modeling process. These considerations have been applied in Sec.~\ref{NG} and \ref{PNM}, respectively.

\subsection{The network generation}
\label{NG}
The pore network modeling consists of two main steps. First, the porous media is mapped into a network of pores connected through throats and, second, simulation of transport processes in the network. In the current work, the bottom-up approach presented in \cite{sadeghi2017pore} is used for generating the hierarchical pore networks. In this approach, the initial network which consists only of nanopores is generated at the first step. Then, some pores are randomly selected and their nearby pores within a certain distance $d_{m}$ are determined. $d_{m}$ is the diameter of the macropore that is inserted at the final step in the empty space created from removing the determined nearby pores. This process repeats until the desired macroporosity ($\varepsilon_{m}$) is achieved. The macroporosity is the metric of the structure hierarchy which is defined as:
\begin{equation}
\nonumber
\varepsilon_{m}=(1- \frac{N}{N_{i}} ) \times 100
\label{c}
\end{equation}
where $N$ is the number of nanopores, and $N_{i}$ is the number of nanopores in the initial nanoporous network. The initial network only consists of nanopores with an unimodal Gaussian distribution of pore diameter about the mean value of $ 20 \; nm$, which is called the average pore size. The pore-to-pore ($ PtoP $) distance of two arbitrary neighboring pores in this initial nanoporous network is $100$ $nm$. Here, the pore size ratio, which is another important metric of the structure hierarchy, can be defined as the ratio of the average diameter of the macropores ($d_{m}$) and that of the nanopores. Addition of macropores to the network which makes the structural hierarchy, changes the initial unimodal pore size distribution and leads to a bimodal pore size distribution. Such an approach for generating hierarchical networks is equivalent to the laboratory synthesis of hierarchical particles with porogens or pore-formers. This process leads to creation of a bimodal form for the pore size distribution which is similar to that of the real laboratory-generated particles. It should be mentioned that, deactivation by coking also changes the network by:
\begin{enumerate}
\item Decreasing the pores and throats' diameter due to coke deposition on the surfaces which is known as the pore narrowing.
\item Plugging of the pores due to the impossibility of crossing reactants from the narrowed pores and throats which is known as the pore blockage.
\end{enumerate}
These changes apply to the network in each time step. It means that the diameter and the blockage of pores and throats should be updated at the end of each time step. Thus, once a hierarchical network is generated, a pore network modeling implements at each time step until reaching the final time which indicates that  a high computational cost is needed for the simulation. Fortunately, it has been shown that the results of 2D simulations obtained by extremely low computational cost are able to describe the quantitative and qualitative behavior of the particle with an acceptable approximation \cite{moghaddam2020investigation}. Therefore, the present study is implemented on 2D networks and consequently, the surface areas and the volumes are calculated accordingly, i.e., the diameter of pore and throat is considered as their cross-section areas, circumference of a pore is considered as its available reaction area, and finally the surface area of the particle is considered as its volume. It should be mentioned that the present study is performed using the open-source OpenPNM libraries written in Python for the pore network modeling of transport phenomena in porous media.

\begin{table}[t]
\caption{List of arameters used in this study.}
    \footnotesize
\renewcommand\arraystretch{1.2}
\begin{tabular}{c c c p{4.1cm}}
\hline\hline
Parameter           & Value & Unit            & Description                                           \\ \hline
    $\langle Da \rangle$& -     & -               & Average pore Damk\"ohler numbers \\
$ N $               & -     & -               & Number of nanopores                                   \\
$ N_{i} $           & -     & -               & Number of nanopores in the initial nanoporous network \\
$ \varepsilon_{m} $ & -     & -               & Macroporosity                                         \\
$ R $               & -     &$    kg/s $      & Main reaction rate                                    \\
$ R_{c} $           & -     &$   kg/m^{2}.s $ & Coking reaction rate                                  \\
$ c $               & -     &$   kg/m^{3} $   & Reactant concentration                                \\
$ c_{c} $           & -     &$   kg/m^{2} $   & Loading of coke per unit pore surface area            \\
 $c_{c}^{max}$&$3\times10^{-6}$&$   kg/m^{2} $   & Maximum loading of coke             \\
$ \rho_{c}  $  & $ 1200 $   &$   kg/m^{3} $   & Density of coke             \\
$ k $               & -     &$   m/s   $      & Main reaction rate constant                           \\
$ k_{c} $           & -     &$   m/s   $      & Reaction rate constant of Coking reaction             \\
$ A $               & -     &$  m^{2}    $    & Active reaction area of pore                          \\
$ a $               & -     & -               & Catalyst activity for main reaction                   \\
$ a_{c} $           & -     & -               & Catalyst activity for coke formation                  \\
$ D $               & -     &$  m^{2}/s $  & Bulk diffusion coefficient of the species             \\
$ D_{e} $           & -     &$   m^{2}/s $  & Effective diffusion coefficient of the species        \\
$ D_{Kn} $          & -     &$   m^{2}/s  $  & Knudsen diffusion coefficient of the species          \\
$ S $               & -     & $ m^{2}  $      & Cross-section area of pore and throat                 \\
$ l $               & -     & $ m $           & Diffusion characteristic length                       \\
$ g $               & -     & $  m^{3}/s $    & Diffusive conductance                                 \\
$ d $               & -     & $ m $           & Pore diameter                                         \\
$ d_{0} $           & -     & $ m $           & Pore diameter at the beginning of coking              \\
$ d_{m} $           &$ 7.5 $& $ nm $          & Molecular diameter of reactant                        \\
$R_{g}$             &$8.314$& $  j/mol.K $    & Universal gas constant                                \\
$T$                &$298.15$& $ K $           & System temperature                                    \\
$M$                 & $ 16 $&$  kg/kmol $     & Molecular weight of the reactant                      \\
$ t $               & -     & $ s $           & Time                                                  \\
\hline\hline
\end{tabular}
\label{nom}
\end{table}

\subsection{Mathematical modeling}
\label{PNM}
Once the network generated and the physical properties are set, modeling of the reaction-diffusion process, and the coking and deactivation can be implemented. The catalyst structure, the diffusion, the reaction, the coking, the deactivation, and their mutual impact and effect on each other are the entities involved in this modeling, which needs a specific approach for investigation as follows:
\begin{enumerate}
\item The concentration profiles are obtained from the pore network modeling of reaction-diffusion.
\item By having concentration distribution, rates of coking and amount of the coke loading are calculated.
\item The catalyst activity is obtained based on the amount of coke deposition.
\item The pores and throats' diameters are recalculated and the pore network structure is updated.
\end{enumerate}

The method and procedure of implementing this approach will be discussed in the following three subsections.

\subsubsection{Reaction-diffusion}
For a given network, the reactive transport process can be mathematically described by the mass balance equation for each pore. For this purpose, only the main reaction and the coke generation reaction are considered and other side reactions are neglected. It should be mentioned that, because the coke generation reaction rate is much slower than the rate of the main reaction \cite{sattler2014catalytic}, the pseudo steady state condition can be considered at each time-step. The mass balance equation for reactant around the pore $ i $ is as follows;
\begin{equation}
\nonumber
\sum_{j=1}^n {g_{i,j} \cdot (c_i - c_j)} - R_{i} = 0 \;\;\;\;\;\;\;\;\;\;\; i=1,2,...,N
\label{balance}
\end{equation}
where $ j $ is the index of the neighboring pores, $ n $ is the number of the neighboring pores, $ N $ is the number of the nanopores, $ g_{i,j} $ is the conductance between pores $ i $ and $ j $, $c_{i}$ is the reactant concentration within pore $ i $, and $R_{i}$ is the main reaction rate within the pore $ i $. For this study, the main reaction is considered as a simple irreversible first-order reaction
\begin{equation}
\nonumber
R_{i}=a_{i}kA_{i}c_{i}\;\;\;\;\;\;\;\;\;\;\; i=1,2,...,N
\label{mainreaction}
\end{equation}
where $a_{i}$ is the catalyst activity for the main reaction within pore $ i $, $ A_{i} $ is the internal surface area of the pore $ i $ and $ k $ is the rate constant. For nonlinear reaction kinetics, nothing changes except the point that the system of equations should be solved iteratively. For a fresh catalyst, the catalyst activity is equal to unity ($ a_{i}=1 $) and gradually decreases due to deactivation that will be discussed in Sec.~\ref{cokdeact}. The diffusive conductance between two neighboring pores reads
\begin{equation}
\nonumber
\frac{1}{ g_{i,j}}=\frac{1}{g_{i}}+\frac{1}{g_t}+\frac{1}{g_{j}}
\label{g}
\end{equation}
with $ g_{i} $ and $ g_{j} $ representing the diffusive conductance of the neighboring pores $ i $ and $ j$, respectively. $ g_{t} $ denotes the diffusive conductance of the connected throat. The diffusive conductance equals
\begin{equation}
\nonumber
g=\frac{D_{e}S}{l}
\label{gl}
\end{equation}
where $l$ is the diffusion characteristic length, $S$ is the cross-section area, and $D_{e}$ is the effective diffusivity which is composed of the effect of the binary and the Knudsen diffusions in parallel. $D_{e}$ can be calculated from
\begin{equation}
\nonumber
\frac{1}{D_{e}}=\frac{1}{D}+\frac{1}{D_{Kn}}
\label{D}
\end{equation}
where $D_{e}$ is the effective diffusivity of the reactant in the mixture, $D$ is the binary diffusivity of the reactant, and $D_{Kn}$ stands for the Knudsen diffusivity that can be expressed as \cite{sadeghi2017pore}
\begin{equation}
\nonumber
D_{Kn}=\frac{ d }{3} \sqrt{\frac{8R_{g}T}{\pi M }}
\label{Kn}
\end{equation}
where $d$ is the pore diameter, $R_{g}$ is the universal gas constant, $T$ is the system temperature, and $M$ is the molecular weight of the reactant. The mentioned relations form a system of equations for computing the concentration for each pore. Also, the total reaction rate can be calculated by summing up the reaction rates in each pore.

\subsubsection{Coking and deactivation}
\label{cokdeact}
The coking rate is directly related to the rate of the main reaction in each pore. Because the first-order kinetics is considered for the main reaction in this study (Eq.\ref{mainreaction}), the coke rate is assumed to have a linear relationship with the reactant concentration as
\begin{equation}
\nonumber
{R_c}_i={a_c}_ik_{c}c_{i}
\label{cokerate}
\end{equation}
where $ {R_c}_i $ is the coking rate in pore $ i$, $ {a_c}_i$ is the catalyst activity for the coke formation in pore $ i$, and $ k_{c} $ is the reaction rate constant of the coking reaction. By assuming that the main reaction and the coke formation reaction occur on the same active sites through a single site mechanism which leads to a uniformly deactivation of these active sites, the catalyst activity for the main reaction ($ a $) and the coke formation reaction ($ a_{c} $) will be equal and can be obtained from \cite{ye2019pore,ye2019optimizing}
\begin{equation}
\nonumber
a=a_{c}=\frac{c_{c}^{max}-c_{c}}{c_{c}^{max}}
\label{deact}
\end{equation}
where $ {c_c}_i $ is the loading of the coke and $ c_{c}^{max} $ is the maximum loading of the coke which completely deactivates the catalyst active sites. Now, the coke loading in each pore at time $ t + \Delta t $ can be calculated from
\begin{equation}
\nonumber
c_{c}(t+\Delta t)=c_{c}(t)+R_{c}(t)\Delta t
\label{coke}
\end{equation}
where $\Delta t $ is the time-step size.

\subsubsection{The network reconstruction}
Deactivation of catalyst by site coverage which is described by Eq.\ref{deact} is one of three mechanisms considered for deactivation. The pore narrowing due to the deposition of coke on the surface of pores is another important mechanism of deactivation. The pore narrowing not only affects the catalyst activity due to reduction of diffusive conductance (see Eq.\ref{gl}), but also the network structure. As time evolves, the pore radius reduction (i.e. pore narrowing) can be determined at each time-step based on the loading of the coke in the pore as
\begin{equation}
\nonumber
\frac{d}{d_{0}}=\sqrt{1-\frac{4c_{c}}{\rho_{c} d_{0}}}
\label{diametercoke}
\end{equation}
where $ \rho_{c} $ is the density of the coke and $ d_{0} $ is the pore diameter with zero loading of the coke (i.e. $ t=0 $ ). Once the pore diameter is obtained at each time-step, the final mechanism of deactivation i.e. the pore blockage can be implemented by comparing the throats diameter with the molecular diameter of reactant ($ d_{m} $). If $ d $ is less than $ d_{m}$,  the pore is considered to be plugged by the coke. It was assumed that no pore blocking occurs at the initial network belonging to the fresh catalyst where there is no trace of coking and deactivation, which means that the minimum pore diameter of the catalyst is equal to or greater than the reactant's molecule diameter. The values of the parameters used in Eqs.~\ref{c}-\ref{diametercoke} are listed in Tab.~\ref{nom}.

\subsubsection{Boundary conditions}
The constant-value boundary condition, i.e. $ c_{b}=1\; mol/m^{3}$ for the external boundary pores has been considered in order to obtain numerical solution of the mathematical model developed for the reactive transport in a hierarchical nanoporous particle.

\section{Results and discussion}
The final aim of this study is to determine a recipe for a catalyst particle that provides the maximum reactivity and high robustness to deactivation with the smallest catalyst material amount. For this purpose, the results are presented in the form of a parametric sweep of mentioned structural properties to specify the effects of different structures on the reactivity and deactivation of the catalyst. It should be mentioned that, in order to make possible such a study of complex system in the parametric form, it is necessary to reduce the computational cost as much as possible by adopting appropriate assumptions, which are listed here:
\begin{enumerate}
\item The catalyst active sites entirely cover the nanopores.
\item The coke uniformly covers the catalyst active sites i.e. the nanopores.
\item The main reaction is considered to have an irreversible first-order kinetics.
\item The coking rate relation with the main reaction is expressed as a linear equation of reactant concentration.
\item The coke generation reaction rate is much slower than the rate of the main reaction.
\item Advection is negligible due to the small length scale of the nanoporous particle.
\item Active sites are only present in nanopores because of negligible contribution of macropores to the total surface area.
\item The minimum pore diameter of the catalyst is equal to or greater than the reactant molecule diameter.
\end{enumerate}

is worth noting that some part or all of these simplifying assumptions can be avoided, and the framework developed here can be used to estimate the particle performance to come up with optimal designs for applications of interest.

\begin{figure}[t]
\centering
\includegraphics[width=\linewidth]{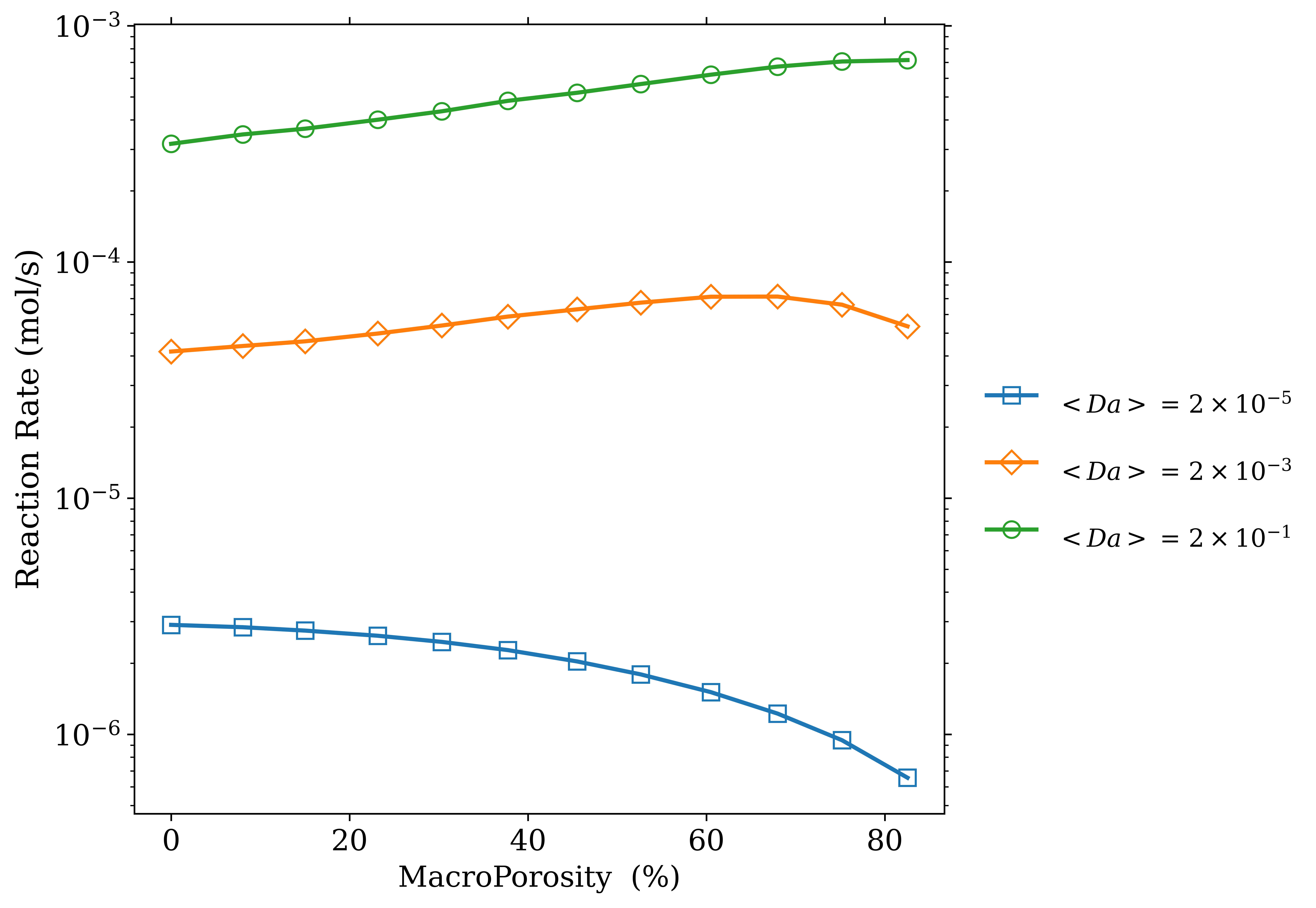}%
\caption{The net reaction rate ($mol/s$) vs. the macroporosity ($ \% $) at different average pore Damk\"ohler numbers; $d_{p} = 40 $ $\mu m $, $ PSR = 20 $.}
\label{Mac}
\end{figure}

\begin{figure*}[t!]
	\centering
	\includegraphics[width=\linewidth]{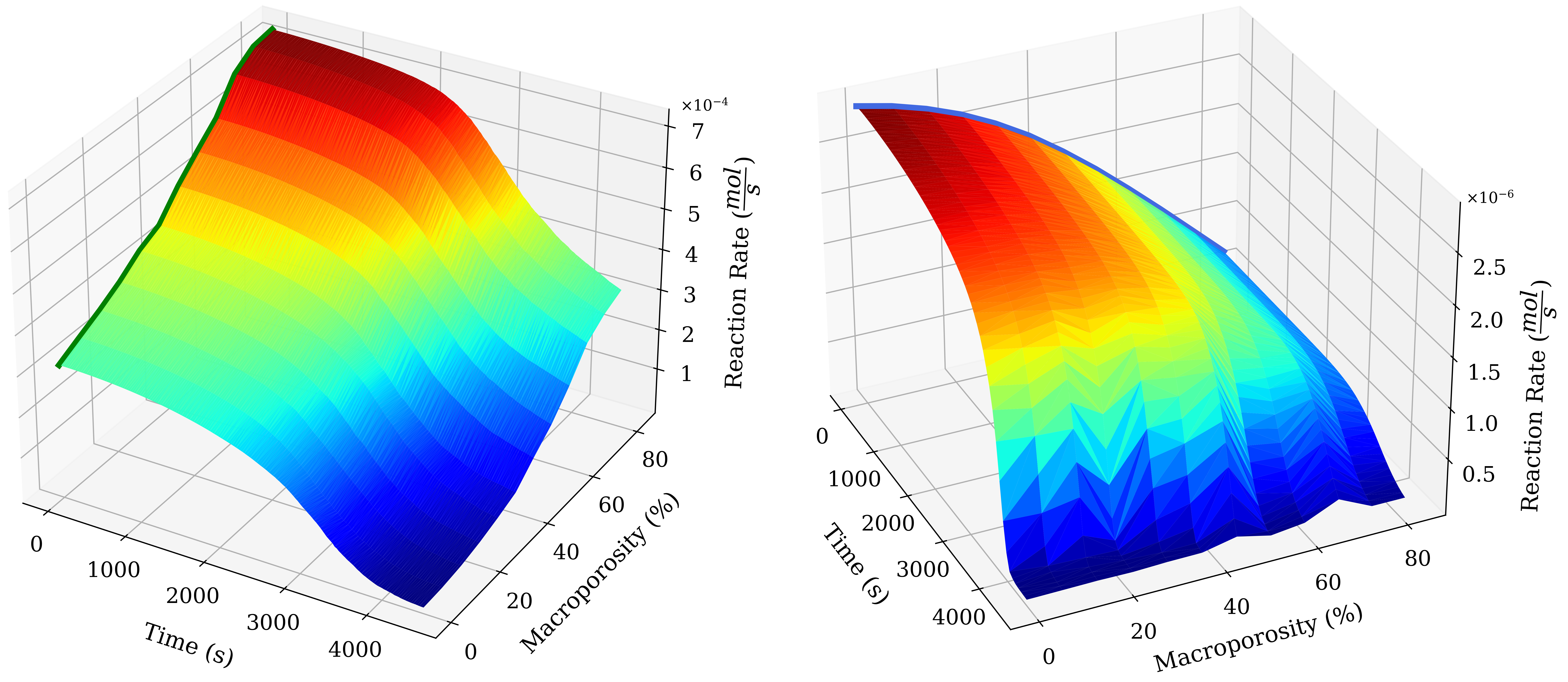}%
	\caption{The net reaction rate ($mol/s$) vs. the macroporosity ($ \% $) and time ($ s $) at high ($ \langle Da \rangle =2\times10^{-1} $, left) and low average pore Damk\"ohler numbers ($ \langle Da \rangle =2\times10^{-5} $, right) ; $d_{p} = 40 $ $\mu m $, $ PSR = 20 $.}
	\label{3d-1-2}
\end{figure*}

\subsection{Effect of macroporosity}
\label{maceff}
In this paper, the pore connectivity ($ Z $) is defined as the number of throats connected to each pore in the initial network and the pore size ratio ($ PSR $) is the ratio of the average diameter of the macropores to that of the nanopores in the hierarchical networks. Fig.~\ref{Mac} shows the performance curves in terms of the net reaction rate ($mol/s$) versus the macroporosity ($ \% $) at different average pore Damk\"ohler numbers within the particle with the diameter of $ 40 $ $ \mu m $, the pore size ratio and the connectivity equal to $ 20 $ and $ 6$, respectively. The average pore Damk\"ohler number $ \langle Da \rangle $ is defined as the average of the local Damk\"ohler numbers in the nanopores \cite{damkohler1937influence},
\begin{equation}
	\nonumber
	\langle Da \rangle =\left\langle \frac{kA_{i}}{ \sum_j   \frac{D_{e}S_{ij}} {l_{ij}} } \right\rangle
	\label{Da}
\end{equation}

The results presented in Fig.~\ref{Mac} are obtained in the absence of deactivation by coking, which can be considered as the initial performance of the fresh catalyst particles. In such a situation, the reactive transport is divided into three conditions; reaction-limited, diffusion-limited, and interstitial. At the reaction-limited condition (i.e. relatively low average pore Damk\"ohler numbers i.e. $ \langle Da \rangle \approx 2\times10^{-5} $), the reactive transport is limited by slow reaction, which means that the diffusion is sufficiently strong, and thus adding macropores reduces the total surface area, leading to a decrease in the net reaction rate. On the contrary, under relatively high average pore Damk\"ohler numbers (i.e. $ \langle Da \rangle \approx 2\times10^{-1} $), the reactive transport is limited by the diffusivity of species. So, adding macropores significantly improves the diffusivity, and particle performance monotonously increases by increasing the macroporosity. Finally, at interstitial condition, a balance between the intensity of the diffusion and reaction is observed at moderate average pore Damk\"ohler numbers (i.e. $ \langle Da \rangle \approx 2\times10^{-5} $). In this condition, adding macropores increases the net reaction rate and the diffusion improvement can outweigh the loss of reaction surface area. From one point onwards, the loss of reactive surface area cannot be compensated by the diffusion improvement and the particle reactivity will start to reduce.

Due to deactivation by coking, the initial performance of the fresh catalysts, presented in Fig.~\ref{Mac}, undergoes changes with time. Fig.~\ref{3d-1-2} presents variation of the net reaction rate versus time and macroporosity at high (left) and low (right) average pore Damk\"ohler numbers. As mentioned above, under relatively high average pore Damk\"ohler numbers ($ \langle Da \rangle =2\times10^{-1} $), the process is diffusion limited and adding macropores improves the transport properties of the particle. Fig.~\ref{3d-1-2}-left reveals that in such a situation, hierarchicality of the particle not only improves its reactivity, but also increases its resistance against deactivation by coking. In the initial times of the deactivation process, the effect of structure hierarchy on the particle performance can be observed as the green line in both Fig.~\ref{Mac} and \ref{3d-1-2}-left, which is monotonously increasing. Fig.~\ref{3d-1-2}-left shows that this trend remains unchanged over time that means that the hierarchical particle has higher coke resistance under relatively high average pore Damk\"ohler numbers. For example, $ 4500 \; s $ after starting the process, a particle with zero macroporosity is completely deactivated while a hierarchical particle with $ 53\% $ of macroporosity has almost $ 50\% $ of its initial activity yet.
	
\begin{figure*}[t!]
	\centering
	\includegraphics[width=0.9\linewidth]{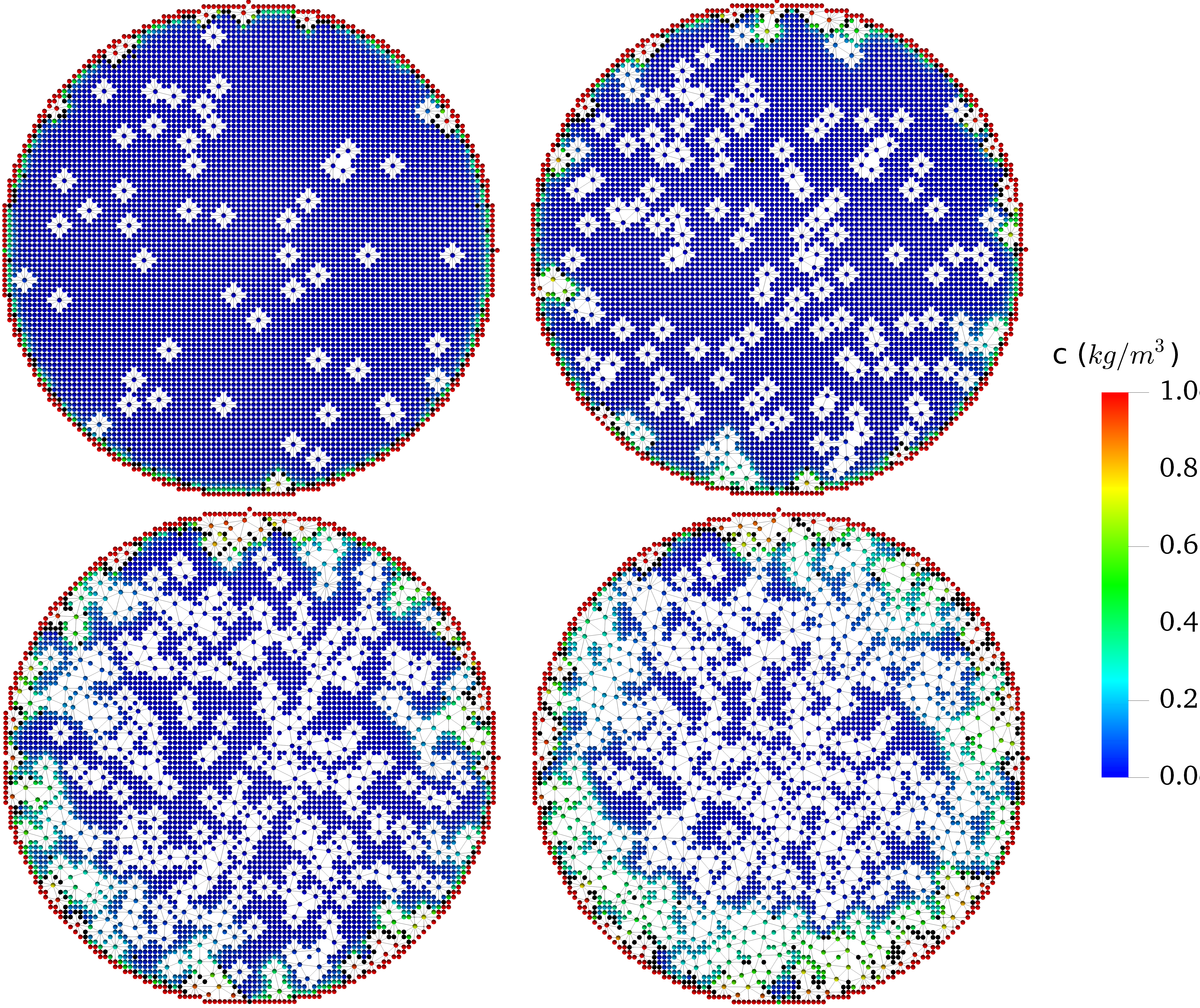}%
	\caption{Concentration contours at high average pore Damk\"ohler numbers ($ \langle Da \rangle =2\times10^{-1} $) after $ 4500\;s $ for four hierarchical porous particles with a diameter $ d_{p} $ of $10 $ $\mu m $, and macroporosities $ \varepsilon_{m} $ of $ 7\% $ (top-left),$  23\% $ (top-right), $ 46\% $ (bottom-left), and $ 60\% $ (bottom-right). Black pores refer to the blocked pores.}
	\label{countours}
\end{figure*}
Higher coke resistance of hierarchical particles at high average pore Damk\"ohler numbers is related to the diffusion path length lengthening by adding macropores. In Fig.~\ref{countours}, the concentration contours are illustrated for four hierarchical particles with different macroporosities. For the sake of clarity, the blocked pores are shown by black circles. This figure shows how macropores facilitate transport into the center of the particle which leads to observation of more uniform concentration distribution in the particle by increasing the macroporosity. The uniformity of concentration distribution, or in other words, lengthening of the diffusion pathways deeper into the particle, not only improves the degree of reactivity but also fortifies the resistance to coking through participating inner pores. At the lowest macroporosity (i.e.  $\varepsilon_{m}=7\%$), only the outer pores are exposed to reaction and consequently deactivation. So by blocking  these outer pores, the particle activity drops to zero as revealed in Fig.~\ref{3d-1-2}. But, by macroporosity increment, for example at $\varepsilon_{m}=60\%$, macropores thwart the effects of pore blockage through the decentralizing of blocked pores from outer regions to the inside. So, unlike the low macroporosity cases, there are no sharp drops in the activity and net reaction rate of the particle decreases with a slighter slope. The key outcome of Fig.~\ref{3d-1-2} is that, at high average pore Damk\"ohler numbers, the pore blockage is confined to outer regions of the particle, and the hierarchical structure propels this phenomenon inwards. Also, a sharp drop in the particle performance is expected at low macroporosities due to this confinedation, which will be analyzed in depth in Sec.~\ref{cmaxeff}.


\begin{figure*}[t]
	\centering
	\includegraphics[width=0.9\linewidth]{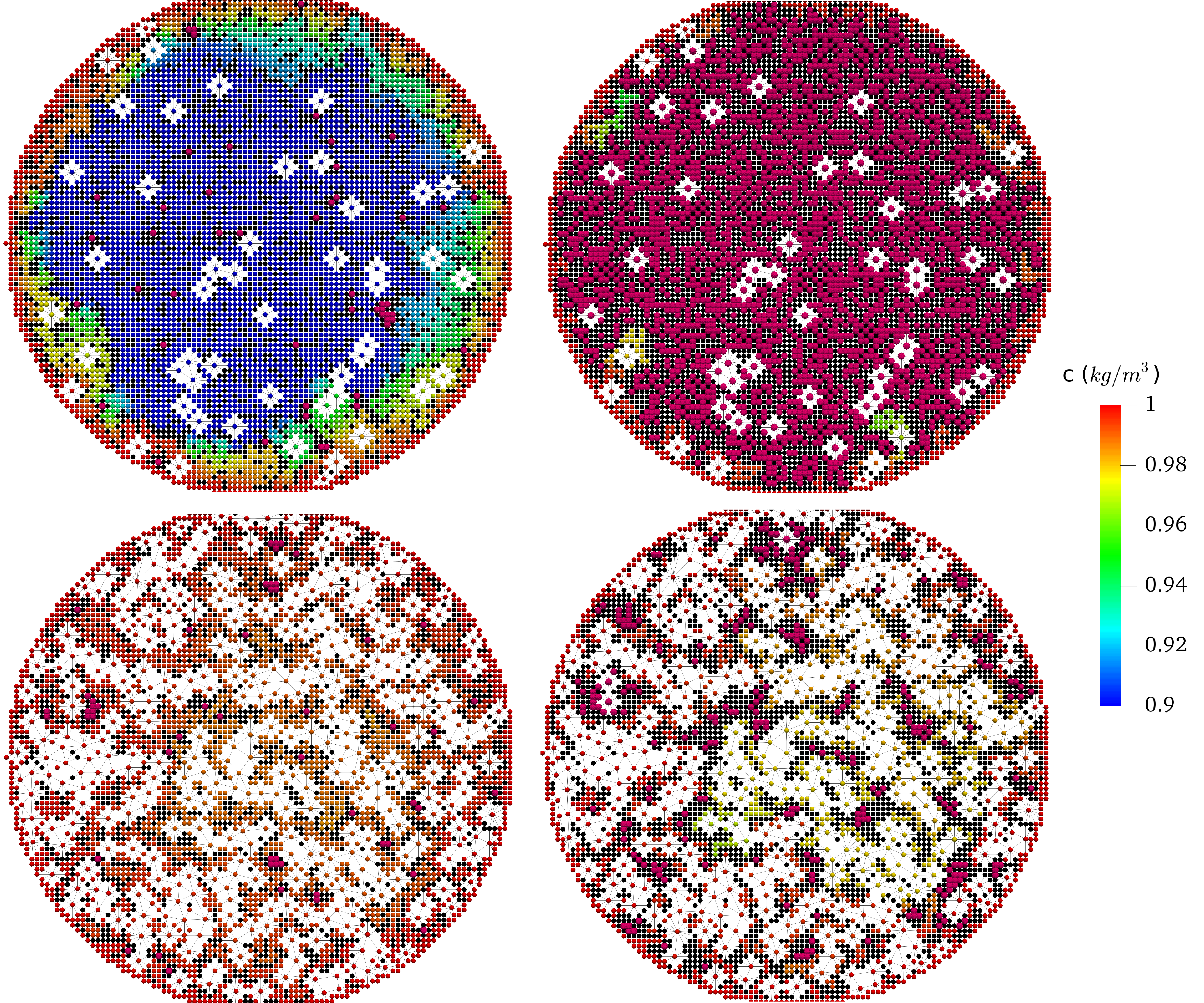}%
	\caption{Concentration contours at low average pore Damk\"ohler numbers ($ \langle Da \rangle =2\times10^{-5} $) after $ 4000\;s $ (top $\&$ bottom-left) and $ 4500\;s $ (top $\&$ bottom-right) for two hierarchical porous particles with a diameter $ d_{p} $ of $10 $ $\mu m $, and macroporosities $ \varepsilon_{m} $ of $ 7\% $ (left $ \& $ right-top) and $ 46\% $ (left $ \& $ right-bottom). Black and pink pores refer to the blocked and inaccessible pores, respectively.}
	
	\label{countours2}
\end{figure*}

\begin{figure*}[t]
	\centering
	\includegraphics[width=\linewidth]{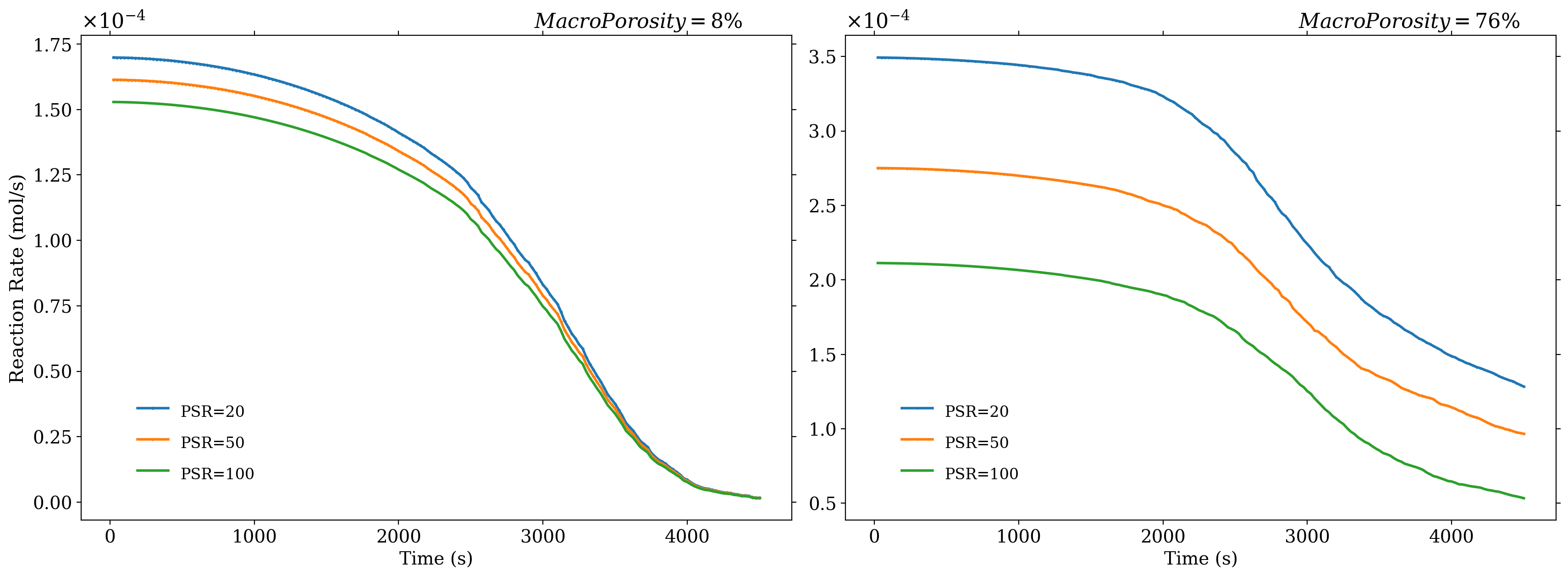}%
	\caption{Net reaction rate ($ mol/s $) vs time ($ s $) at different pore size ratios and macroporosities ($ \% $) under high average pore Damk\"ohler number ($ \langle Da \rangle =2\times10^{-1} $);  $d_{p} = 40 $ $\mu m $, $ PSR = 20 $.}
	
	\label{PSR}
\end{figure*}


Fig.~\ref{3d-1-2}-right reveals the temporal changes in the catalyst reactivity at low average pore Damk\"ohler number ($ \langle Da \rangle =2\times10^{-5}$), which is known as reaction limited process. As mentioned above, in such a situation, the diffusion rate is far superior to reaction rate, so adding macropores has a little effect on transport properties, but decreases the total surface area. Fig.~\ref{3d-1-2}-right reveals that the particle hierarchicality not only reduces its reactivity, but also decreases its resistance against deactivation by coking. Effect of macroporosity on the particle reactivity at the beginning times of the deactivation process, can be observed in the monotonously decreasing blue line shown in both Figs.~\ref{Mac} and \ref{3d-1-2}-right. This monotonously decreasing trend is kept in almost all time steps illustrated in Fig.~\ref{3d-1-2}, which means that adding macropores has reduced the catalyst coke resistance for the reaction limited process (i.e. low average pore Damk\"ohler numbers). In the reaction limited process, diffusion rate of the species is sufficiently strong which leads to the uniform distribution of concentration inside the particle spatially at the initial time steps. In such a situation, the blocked pores begin to germinate everywhere inside the particle. By increasing the number of plugged pores, it becomes possible for some pores to be surrounded by plugged pores and become unavailable. It is clear that these inaccessible pores lead to a decreasing reaction rate that seems to be compensated by adding macropores. Fig.~\ref{countours2} shows the concentration contours for two hierarchical particles with different macroporosities at different time-steps. In this figure, the blocked and inaccessible pores are shown by black and pink pores, respectively. A comparison between Fig.~\ref{3d-1-2}-right and \ref{countours2} reveals that  adding macropores have reduced the number of inaccessible pores, but in return has reduced the total surface area. So, it can be concluded that the hierarchical and non-hierarchical particles have almost the same deactivation time while the latter is some order of magnitudes more active than the former. This means that the hierarchical catalysts are not always favorable and their advantage is solely for systems where there is no trace of the reaction limited process.

In the case of non-hierarchical particle ($macroporosity=0$), a comparison between the particle reactivity and deactivation (presented in Fig.~\ref{3d-1-2}) and its equivalent concentration contours (presented in Figs.~\ref{countours} and ~\ref{countours}) reveals that inaccessibility of pores due to blocking of their surrounded pores is the major reason of particle deactivation. But, there is a difference between occurrences of this phenomenon at different average pore Damk\"ohler numbers. Under relatively high average pore Damk\"ohler number, the internal pores are naturally inaccessible because of hindered diffusivity. So, their inaccessibility due to being surrounded by blocked pores has less impact on the particle reactivity in comparison to inaccessibility of the internal pores of particles with low average pore Damk\"ohler number. This is why at $ \langle Da \rangle =2\times10^{-1}$, the reactivity loss of the particle has a maximum rate of $ 1.75 \times 10^{-7}\; mol/s^{2} $ while this maximum rate at $\langle Da \rangle =2\times10^{-5} $ is almost $ 20 $ times higher ($ 3.2 \times 10^{-6}\; mol/s^{2} $).




\subsection{Effect of the pore size ratio}
The pore size ratio is an important structural parameter in design and application of the hierarchical catalysts. Existence of bi- or multi-modal form in the pore size distribution of the hierarchical materials is the result of the pore size ratio considered in their design. Fig.~\ref{PSR} shows the net reaction rate over time for two particles with low and high macroporosities under relatively high average pore Damk\"ohler number ($ \langle Da \rangle =2\times10^{-1} $) at different pore size ratios and macroporosities. In this figure, the same macroporosity is created at different pore size ratios, i.e., there is a larger number of macropores in the particle with smaller pore size ratio. This means that the macropores' distribution inside the particle with smaller pore size ratio is more uniform which leads to uniformity of concentration distribution.

So, as shown in Fig.~\ref{PSR}, at high average pore Damk\"ohler numbers where the process is diffusion limited, the particle with smaller pore size ratio is more reactive due to its uniform distribution of smaller macropores which increases the diffusion pathways deeper into the particle and improves the reactivity. Fig.~\ref{PSR} also reveals that at low macroporosities, the deactivation resistance of particles with different pore size ratios are almost the same. But, as the macroporosity increases by increasing the number of macropores, particles with smaller pore size ratios show better catalyst protection against the deactivation. This is because of the uniform distribution of small macropores which reduces the inaccessibility of inner pores and provides their better participation in the reaction and deactivation process.

\subsection{Effects of catalyst coking capacity}
\label{cmaxeff}
The maximum loading of the coke that completely deactivates the catalyst's active sites is $ c_{c}^{max}$ which is independent of the particle microstructure. Based on Eq.~\ref{deact}, the value of the coke in a pore is always equal to or less than $ c_{c}^{max}$, because when the coke content of a pore reaches this maximum value, its activity becomes zero and no more coke formation occurs there. But, it should be mentioned that the real coke capacity of a catalyst can not be considered as the sum of $ c_{c}^{max}$ for all pores because it is an ideal case that its occurrence depends on the network structure and the reaction condition. The capacity for a catalyst particle to accommodate coke can be determined by the dimensionless coke content, which is the ratio of the actual coke content to the maximum coke content where all pores have $ c_{c}^{max}$ value of the coke \cite{biswas1987unified}.

\begin{figure}
\centering
\includegraphics[width=\linewidth]{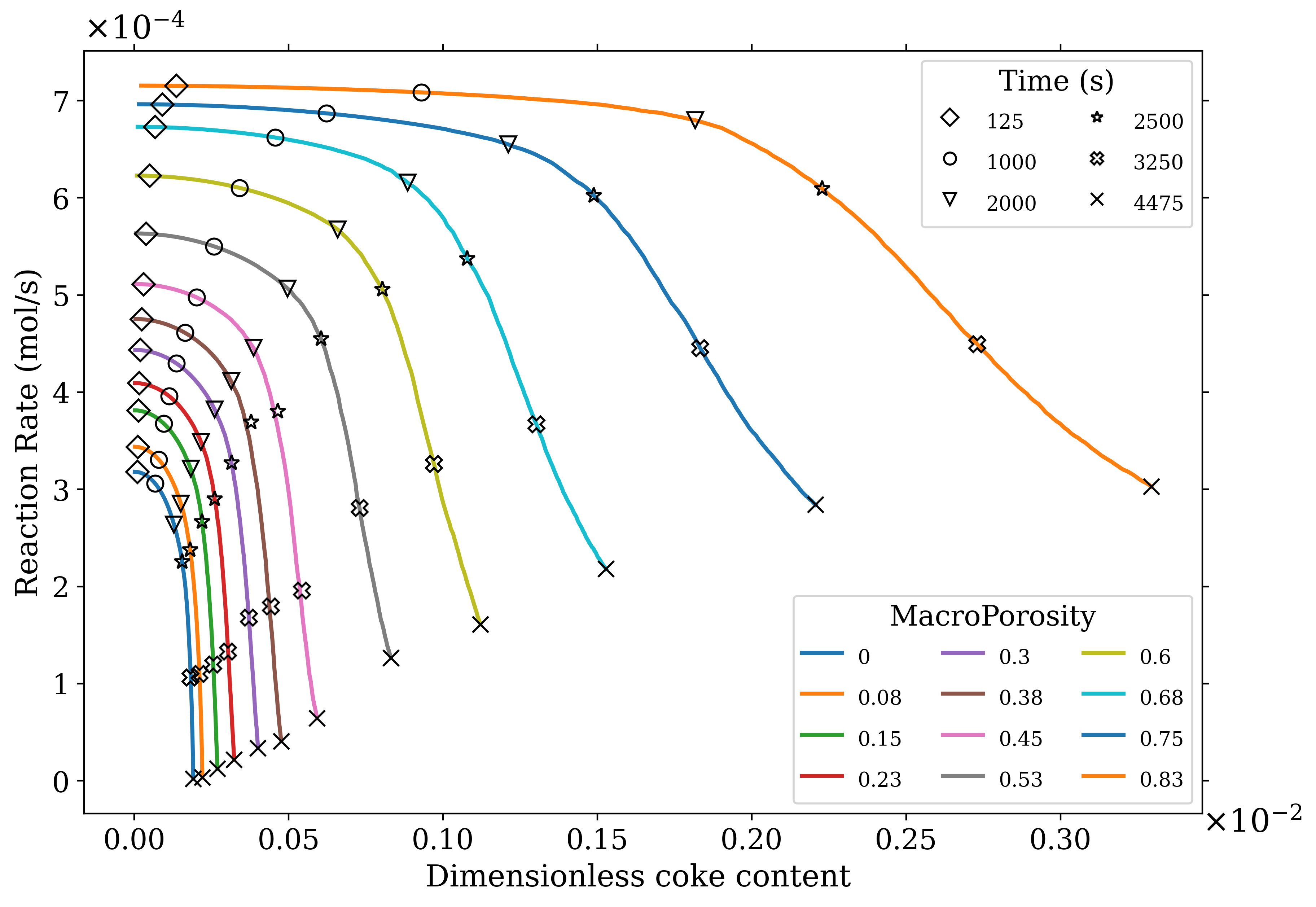}%
\caption{Net reaction rate ($ mol/s $) vs. dimensionless coke content at different macroporosities ($ \% $) under high average pore Damk\"ohler number ($ \langle Da \rangle =2\times10^{-1} $);  $d_{p} = 40 $ $\mu m $, $ PSR = 20 $.}

\label{DCC}
\end{figure}

Fig.~\ref{DCC} shows the net reaction rate ($ mol/s $) against the dimensionless coke content at different macroporosities for a particle with a radius of $20 $ $\mu m $ and an average pore size ratio of $ 20$. These results have been obtained in a time interval of $ 4500 \; s $ and some of these times are marked out. Fig.~\ref{DCC} reveals that by increasing time (spatially at low macroporosities), a sharp drop occurs in the activity of the catalyst which is due to increasingly growth in the number of blocked and inaccessible pores which was mentioned in Sec.~\ref{maceff}. By considering this critical time, the deactivation of the catalyst can be divided into two conditions that we name them as kinetical deactivation and structural deactivation. At the kinetical deactivation condition, the deactivation occurs due to the reduction of the catalyst activity presented in Eq.~\ref{deact}. In such a situation, the effects of pore blockage and inaccessibility are not dominant and if the process runs under this condition, its dimensionless coke content will reach $ 1 $ when its net reaction rate is reduced to zero during a long time interval. But, due to occurrence of the pore blockage, and consequently the pore inaccessibility, which can be considered as a modification in the catalyst structure, the deactivation process becomes more intensified. At this structural deactivation condition, many pores are no longer available for the reactants that leads to a sharp drop in the catalyst performance.

\begin{figure}
\centering
\includegraphics[width=\linewidth]{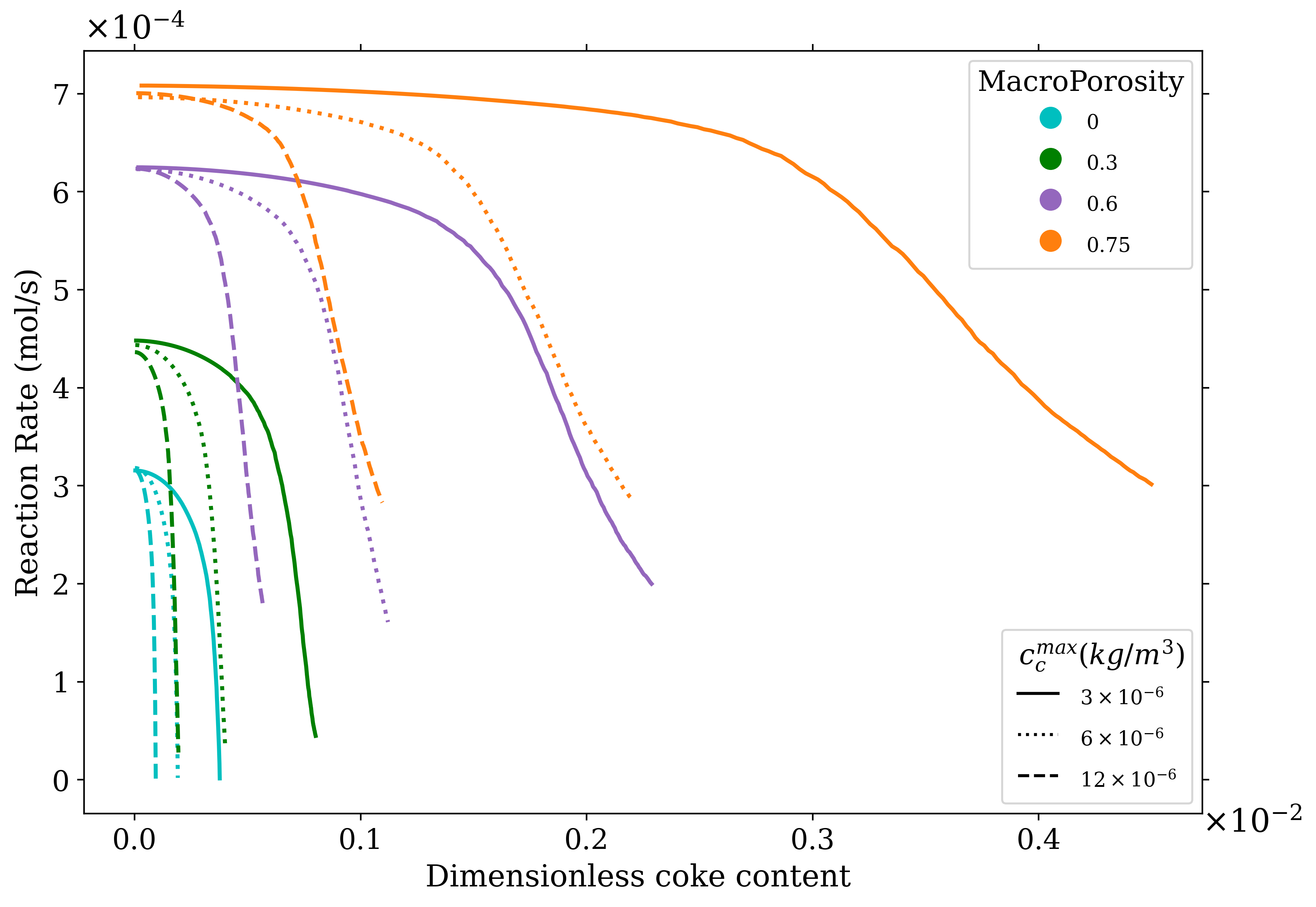}%
\caption{Net reaction rate ($ mol/s $) vs. time ($ s $) at different $ c_{c}^{max}$ ($ kg/m^{2} $) and macroporosities ($ \% $) under high average pore Damk\"ohler number ($ \langle Da \rangle =2\times10^{-1} $);  $d_{p} = 40 $ $\mu m $.}

\label{cmax}
\end{figure}

Fig.~\ref{cmax} reveals the effects of the catalyst $ c_{c}^{max}$ on its reactivity and deactivation. When the value of $ c_{c}^{max}$ is relatively small, the deactivation of pores (Eq.~\ref{deact}) occurs before its blocking. In such a situation, the pore reactivity (and consequently the coke formation) becomes zero but its participation in mass transport remains operative. However, in the case of relatively large $c_{c}^{max}$ values, the pore blockage happens when the pore is still active and stops the occurrence of the reaction as well as the diffusion in that pore. So, it is expected that a catalyst with larger values of $ c_{c}^{max}$ underwent a much more severe deactivation. Fig.~\ref{cmax} reveals that by increasing the value of $ c_{c}^{max}$, the kinetical deactivation condition turns into structural deactivation condition which, as discussed above, results in a sharp drop in the catalyst performance.

In Sec.~\ref{Introduction}, it has explained that most of the effort to design deactivation resistant catalysts is dedicated to design of active sites which leads to achieving catalysts with different values of $ c_{c}^{max}$. Fig.~\ref{cmax} presents the effects of these efforts combined with the effect of microstructure modifications on the reactivity and deactivation of the catalyst. Based on these results, at low macroporosities the particles with different values of $ c_{c}^{max}$ undergo a severe structural deactivation condition but as the macroporosity increases, this deactivation condition becomes more kinetical. This is an important outcome about the performance of hierarchical porous catalyst particles for the rational design of robust catalysts against deactivation by coking.

\section{Conclusion}
A novel pore network model has been presented to model the pore-scale reaction-diffusion processes combined with the deactivation and coking in a hierarchical porous particle. Three deactivation mechanisms considered including site coverage, pore narrowing, and pore blockage. The hierarchical microstructure created through adding macropores inside the initial nanoporous network and interactions between the obtained hierarchical network structure (i.e. the macroporosity and the size ratio of macropores to nanopores) and the reaction-diffusion processes were investigated under different reaction conditions (i.e. average pore Damk\"ohler numbers) and deactivation (i.e. maximum coke loading of the catalyst).

The results showed that the deactivation of the particle mainly depends on the average pore Damk\"ohler number. At relatively high average pore Damk\"ohler numbers, the pore blocking occurs near the particle boundaries while under relatively low average pore Damk\"ohler numbers, the pore blockage and inaccessibility is seen everywhere inside the particle. Thus, it is concluded that depending on the Damk\"ohler number, increasing the macroporosity does not necessarily improve the performance of the particle.

In summary, at high average pore Damk\"ohler number, the hierarchical particles have higher reactivity and higher coke resistance while under low average pore Damk\"ohler number, the hierarchicality reduces the particle reactivity but does not affect its coke resistance. Also, it is found that depending on the maximum coke loading of the catalyst, two distinct trends can be observed for the particle deactivation by time increment, including a sharp drop, and a slightly decrease at high, and low maximum coke loading of the catalyst, respectively. It is found that at high average pore Damk\"ohler numbers, increasing the particle macroporosity and the pore size ratio, and reducing the catalyst maximum coke loading can significantly improve its reactivity and deactivation resistance. The results obtained in this study, can be used to guide the rational design of hierarchical catalyst with improved reactivity and more robust against deactivation by the coke formation.

\section*{Acknowledgment}
Special acknowledgment goes to Dr. Jeff T. Gostick and Dr. Mohammad Amin Sadeghi from university of Waterloo to their technical support about OpenPNM. We also are grateful to Dr. Abdolah Sepahvand from the department of computer engineering $\&$ information technology, Amirkabir university of technology to provide computational requirements of the present study. Additionally, sincerest appreciation should also be extended to Dr. Hamid Ghasabzadeh from catalysis $\&$ nanotechnology research division of Research Institute of Petroleum Industry (RIPI) for his warm discussions on the hierarchical catalyst microstructure and performance.

\section*{References}
\bibliographystyle{elsarticle-num}
\bibliography{citations}

\begin{thebibliography}{10}
\expandafter\ifx\csname url\endcsname\relax
  \def\url#1{\texttt{#1}}\fi
\expandafter\ifx\csname urlprefix\endcsname\relax\def\urlprefix{URL }\fi
\expandafter\ifx\csname href\endcsname\relax
  \def\href#1#2{#2} \def\path#1{#1}\fi

\bibitem{sadeghi2017pore}
M.~A. Sadeghi, M.~Aghighi, J.~Barralet, J.~T. Gostick, Pore network modeling of
  reaction-diffusion in hierarchical porous particles: The effects of
  microstructure, Chemical Engineering Journal 330 (2017) 1002--1011.

\bibitem{chen2018pore}
L.~Chen, R.~Zhang, T.~Min, Q.~Kang, W.~Tao, Pore-scale study of effects of
  macroscopic pores and their distributions on reactive transport in
  hierarchical porous media, Chemical Engineering Journal 349 (2018) 428--437.

\bibitem{moghaddam2020investigation}
M.~Moghaddam, A.~Abbassi, J.~Ghazanfarian, S.~Jalilian, Investigation of
  microstructure effects on performance of hierarchically structured porous
  catalyst using a novel pore network model, Chemical Engineering Journal 388
  (2020) 124261.

\bibitem{dashtian2019convection}
H.~Dashtian, S.~Bakhshian, S.~Hajirezaie, J.-P. Nicot, S.~A. Hosseini,
  Convection-diffusion-reaction of co2-enriched brine in porous media: A
  pore-scale study, Computers \& Geosciences 125 (2019) 19--29.

\bibitem{vovcka2000pore}
R.~Vo{\v{c}}ka, M.~A. Dubois, Pore network as a model of porous media:
  Comparison between nonhierarchical and hierarchical organizations of pores,
  Physical Review E 62~(4) (2000) 5216.

\bibitem{petropoulos1991network}
J.~H. Petropoulos, J.~K. Petrou, A.~I. Liapis, Network model investigation of
  gas transport in bidisperse porous adsorbents, Industrial \& Engineering
  Chemistry Research 30~(6) (1991) 1281--1289.

\bibitem{meyers1998network}
J.~Meyers, A.~Liapis, Network modeling of the intraparticle convection and
  diffusion of molecules in porous particles packed in a chromatographic
  column, Journal of Chromatography A 827~(2) (1998) 197--213.

\bibitem{meyers1999network}
J.~Meyers, A.~Liapis, Network modeling of the convective flow and diffusion of
  molecules adsorbing in monoliths and in porous particles packed in a
  chromatographic column, Journal of Chromatography A 852~(1) (1999) 3--23.

\bibitem{wang2021pore}
S.~Wang, X.~Yang, Y.~He, Pore-scale study of reactive transfer process
  involving coke deposition via lattice boltzmann method, AIChE Journal (2021)
  e17478.

\bibitem{lin2021lattice}
Y.~Lin, C.~Yang, C.~Choi, W.~Zhang, H.~Machida, K.~Norinaga, Lattice boltzmann
  simulation of multicomponent reaction-diffusion and coke formation in a
  catalyst with hierarchical pore structure for dry reforming of methane,
  Chemical Engineering Science 229 (2021) 116105.

\bibitem{ye2019optimizing}
G.~Ye, H.~Wang, X.~Zhou, F.~J. Keil, M.-O. Coppens, W.~Yuan, Optimizing
  catalyst pore network structure in the presence of deactivation by coking,
  AIChE Journal 65~(10) (2019) e16687.

\bibitem{rao2012increasing}
S.~M. Rao, M.-O. Coppens, Increasing robustness against deactivation of
  nanoporous catalysts by introducing an optimized hierarchical pore
  network—application to hydrodemetalation, Chemical Engineering Science 83
  (2012) 66--76.

\bibitem{schmidt2013coke}
F.~Schmidt, C.~Hoffmann, F.~Giordanino, S.~Bordiga, P.~Simon,
  W.~Carrillo-Cabrera, S.~Kaskel, Coke location in microporous and hierarchical
  zsm-5 and the impact on the mth reaction, Journal of Catalysis 307 (2013)
  238--245.

\bibitem{froment2001modeling}
G.~Froment, Modeling of catalyst deactivation, Applied Catalysis A: General
  212~(1-2) (2001) 117--128.

\bibitem{arbabi1991computer}
S.~Arbabi, M.~Sahimi, Computer simulations of catalyst deactivation—i. model
  formulation and validation, Chemical Engineering Science 46~(7) (1991)
  1739--1747.

\bibitem{sattler2014catalytic}
J.~J. Sattler, J.~Ruiz-Martinez, E.~Santillan-Jimenez, B.~M. Weckhuysen,
  Catalytic dehydrogenation of light alkanes on metals and metal oxides,
  Chemical Reviews 114~(20) (2014) 10613--10653.

\bibitem{ye2019pore}
G.~Ye, H.~Wang, X.~Duan, Z.~Sui, X.~Zhou, M.-O. Coppens, W.~Yuan, Pore network
  modeling of catalyst deactivation by coking, from single site to particle,
  during propane dehydrogenation, AIChE Journal 65~(1) (2019) 140--150.

\bibitem{damkohler1937influence}
G.~Damk{\"o}hler, Influence of diffusion, fluid flow, and heat transport on the
  yield in chemical reactors, Der Chemie Ingenieur 3 (1937) 359--485.

\bibitem{biswas1987unified}
J.~Biswas, D.~Do, A unified theory of coking deactivation in a catalyst pellet,
  The Chemical Engineering Journal 36~(3) (1987) 175--191.

\end{thebibliography}
\end{document}